\documentclass[
 reprint,
nofootinbib,
 amsmath,amssymb,
 aps,
]{revtex4-2}

\usepackage{graphicx}
\usepackage{dcolumn}
\usepackage{bm}
\usepackage{float}
\usepackage{xcolor}
\usepackage{hyperref}

\makeatletter
\newcommand{\leqnomode}{\tagsleft@true\let\veqno\@@leqno}
\newcommand{\reqnomode}{\tagsleft@false\let\veqno\@@eqno}
\makeatother

\begin{document}

\preprint{APS/123-QED}

\title{Diffusion approximation of a network model of meme popularity}

\author{Kleber A.~Oliveira\ref{corresp}}
\author{Samuel Unicomb}
\author{James P.~Gleeson}
\affiliation{MACSI, Department of Mathematics and Statistics, University of Limerick, Limerick, Ireland}

\begin{abstract}
Models of meme propagation on social networks, in which memes compete for limited user attention, can successfully reproduce the heavy-tailed popularity distributions observed in online settings. While system-wide popularity distributions have been derived analytically, the dynamics of individual meme trajectories have thus far evaded description. To address this, we formulate the diffusion of a given meme as a one-dimensional stochastic process, whose fluctuations  result from aggregating local network dynamics using classic and generalised central limit theorems, with the latter based on stable distribution theory. Ultimately, our approach decouples competing trajectories of meme popularities, allowing them to be simulated independently, and thus parallelised, and expressed in terms of Fokker-Planck equations.
\end{abstract}

\maketitle

\section{Introduction}\label{sec:introduction}

The dynamics of information diffusion in networked systems have been investigated extensively in recent years~\cite{zhang2016dynamics}. Broadly interpreted as transmissible pieces of information~\cite{ferrara2013clustering}, memes have been collected and analysed in a number of online empirical studies~\cite{hodas2014simple, cheng2014cascades, miotto2017stochastic, notarmuzi2022universality}. Of particular interest is meme popularity, which corresponds to the number of times that a given piece of information content has been shared by a social media user to their network neighbours, or followers. Empirically, distributions of meme popularity are characterised by their high variance, and indeed heavy tails. That is, while most memes die out soon after being introduced in favour of competing content, a small number are heavily shared, both through viral cascades and the broadcasting effect of nodes with high numbers of followers~\cite{goel2016structural}. 

The distributions of meme popularity observed in empirical settings have been recovered in simple models, through a mechanism known as competition-induced criticality, or CIC~\cite{gleeson2014competition}. In these models, memes are viewed as being in competition for the finite resource that is user attention. The constraint of finite attention has become key in many models of social network dynamics~\cite{weng2012competition, ciampaglia2015production, lorenz-spreen2019accelerating, lerman2016information}. In models of CIC, this constraint is typically represented by a screen attributed to each user, on which a single meme can be held; extensions to this model can allow a list of memes~\cite{gleeson2016effects}. Memes are transmitted when users broadcast to their followers the contents of their own screen, overwriting any content that was previously on the screens of the followers. Typical of critical phenomena on complex networks~\cite{dorogovtsev2008critical}, dynamic quantities here can be described in terms of branching processes.

This approach to describe the phenomenon is also known as a tree-like approximation, in the sense it assumes cascade growth outcomes typically neglect the presence of cycles in the network structure -- a useful assumption to a very reasonable degree~\cite{chandra2020critical}. Additionally, the CIC models can be understood as neutral models~\cite{bentley2021neutral, pinto2011quasi}, meaning no meme has any inherent characteristic to help them grow more popular than others; in other words, the popularity dynamics are solely controlled by macroparameters such as the underlying network structural characteristics and the rate new memes are introduced in the system, while trajectories of individual memes are highly stochastic.

Although branching process techniques have been successful in deriving system-wide distributions of meme popularity, they are poorly suited to describing the time evolution of the popularity of any single meme. In simulations of the CIC dynamics, the popularity of a given meme evolves according to a discrete-time jump process, where the number of screens that it occupies increases and decreases due to competition with other memes for user attention. We examine an approximation that considers the fraction of screens occupied by memes to be a continuous variable.

Viewing meme popularity as a continuous, stochastic quantity suggests connection to related models. For example, Di Santo~\textit{et al}.~\cite{disanto2017simple} encode the statistics of several avalanche types through the formalism of stochastic equations. Notarmuzi~\textit{et al}.~\cite{notarmuzi2018analytical} also produce a diffusion process description for the quality-biased competition model (which is inspired by the CIC model) that provides novel analytical understanding of the mechanisms prescribed by agent-based simulations. In this paper, we carefully examine the discrete-to-continuous limit to produce novel descriptions of meme dynamics making use of the classic or generalised central limit theorems according to certain cases, and examine the accuracy and parameter dependence of the resulting approximations.

\begin{table*}
\caption{Potential outcomes of an update step in the model dynamics. Case is determined by whether a user selected uniformly at random innovates new content, which occurs with probability $\mu$, and whether their screen contains the red meme upon selection, which occurs with probability $s$. The probability of case $j$ occurring is $\pi_j$. Probability generating functions $g_j(x)$ describe the fluctuations in $s$ when case $j$ arises, with $f(x)$ the generating function of the out-degree distribution. \label{tab:table1}}
\begin{ruledtabular}
\begin{tabular}{llllll}
    case $j$ & innovates & red meme selected & contribution & $\pi_j$ & $g_j(x)$ \\
    \hline \hline
    1 & yes & yes & $-$ & $\mu s$             & $1 - \pi_1 + \pi_1 x f(1 - s + s x)$  \\[1ex]
    2 & yes & no  & $-$ & $\mu (1 - s)$       & $1 - \pi_2 + \pi_2 f(1 - s + s x)$    \\[1ex]
    3 & no  & yes & $+$ & $(1 - \mu) s$       & $1 - \pi_3 + \pi_3 f(s + (1 - s) x)$  \\[1ex]
    4 & no  & no  & $-$ & $(1 - \mu) (1 - s)$ & $1 - \pi_4 + \pi_4 f(1 - s + s x)$
\end{tabular}
\end{ruledtabular}
\end{table*}

The remainder of this paper is organised as follows. In Section~\ref{sec:diffusionapprox}, we begin by describing the competition-induced criticality 
model, as well as the simulation scheme used in numerical experiments. Then in Section~\ref{sec:fluctuations}, we identify distinct cases of local fluctuations that can be aggregated by means of central limit theorems. In Sections~\ref{sec:finite} and \ref{sec:infinite}, we distinguish between the case of finite- and infinite-variance degree distributions, making use of classic and generalised central limit theorems to derive stochastic, finite-difference equations modelling popularity dynamics. In Section~\ref{sec:results} we validate our analysis, by systematically comparing the results of our one-dimensional model to those obtained through computational simulations of the CIC model on various networks.

\section{The diffusion approximation}\label{sec:diffusionapprox}

In this section we describe a diffusion approximation of the meme popularity model of Ref.~\cite{gleeson2014competition}. Following that work, we assume a directed network of size $N$, with edge direction representing the direction of the flow of information in a social network. As such, the out-degree $k$ of a node is the number of followers, for instance, in an online social network like Twitter. We assume this to follow a distribution $p_k$. The average degree is $\langle k \rangle$, and its variance is $\sigma^2$. For simplicity, we assume that $\langle k \rangle$ is integer and that the in-degree distribution is $\langle k \rangle$-regular, and wired according to the configuration model, meaning the network is maximally random up to in- and out-degree. Each node represents a screen on which a single meme can be displayed. Each meme has a discrete type, or colour, and is copied from the screen of a given node to its followers when the node ``tweets.''

In each discrete time step, the system is updated by having one node is selected uniformly at random to tweet. With probability $\mu$, the active node innovates and tweets an original meme, overwriting the previously held meme, as well as those of its out-neighbours. Otherwise, i.e., with probability $1 - \mu$, the node ``retweets'' the meme currently on its screen and so overwrites the screens of its out-neighbours with the retweeted meme. We define the elementary time unit as $dt = 1 / N$ between successive updates, such that a macroscopic time step of $1$ is when $N$ updates have occurred across the system. We initialise the system by assuming every screen to be blank, and assume that if a blank screen is chosen for update, it innovates a new meme with probability $1$. This model was studied in Ref.~\cite{gleeson2014competition} using a branching process approximation for the distribution of meme popularity (accumulated number of retweets). In this paper, we focus on characterising the stochastic trajectories of individual memes, which necessitates a discrete-to-continuous approximation to study the fraction of screens occupied by a meme at a given time.

\subsection{Fluctuation cases}\label{sec:fluctuations}

We focus on a particular meme, which we shall refer to as the \emph{red} meme, that occupies a fraction $s$ of all screens in the system. The evolution of this meme is a stochastic function of the network topology $p_k$, the innovation rate $\mu$, as well as the current density $s$ of the meme itself. The temporal evolution of $s$ can be modelled by identifying four cases, or outcomes, of an update step. These are indexed by $j = 1, 2, 3$ and $4$, as outlined in Table~\ref{tab:table1}. In cases 1 and 2, the selected node innovates original content, and in cases 3 and 4 it simply retweets. In cases 1 and 3, the selected node contains the red meme, and in cases 2 and 4, some other meme. Note that cases 1, 2 and 4 can only diminish the number of red memes on screens across the system, by overwriting with some other meme. In contrast, case 3 can only increase the number of red memes, by overwriting the screens of its followers with the red meme. The sign of the contribution and the probability $\pi_j$ of case $j$ arising are given in columns 4 and 5 of Table~\ref{tab:table1}. Further, upon the assumption that red memes are distributed with density $s$ randomly amongst all nodes, we can model the fluctuation in $s$ at a given step, in each case $j$. The probability generating function, or pgf, corresponding to the out-degree distribution $p_k$ is defined to be $f(x)=\sum_k p_k x^k$. Accordingly, the pgf of the number of neighbours that do or do not contain meme $s$ is given by composition of $f$ with the pgf of the Bernoulli distribution: $1 - s + sx$ or $s + (1 - s)x$, respectively. By shifting and scaling by the probabilities $\pi_j$, we define in the final column of Table~\ref{tab:table1} the pgfs $g_j(x)$ for the integer change in the absolute number of nodes containing the red meme due to each case $j$. 
The goal of a diffusion approximation~\cite{nolan2020univariate} is to aggregate a large number of fluctuations resulting from microscopic dynamics to estimate an effective change in the variable of interest over a larger time interval. This is achieved using central limit theorems, which deal with sums of random variables. In this way, we approximate the temporal evolution of $s$, using our knowledge of the fluctuations in $s$ due to individual updates, as described by the distributions $g_j(x)$. We group together $n$ successive update events, and estimate the resulting change to the red meme, $\Delta s$. The time step in the diffusion approximation is $\Delta t = n\, dt = n / N$. In choosing a value for $n$, we weigh two considerations. The first is that $n$ must be sufficiently large for the sum of fluctuations to be well-approximated using the central limit theorem. The second is that $n$ must not be so large that the value of $s$ varies appreciably over the interval $\Delta t$, since our approximation assumes fluctuations to be independent, meaning $s$ is assumed constant over the $n$ fluctuations.

\begin{table*}
\caption{Properties of the generating functions $g_j(x)$ required by the central limit theorems. The mean and variance of $C_j$ are used in the classical central limit theorem, and the tail parameters $c_j$ in the generalised central limit theorem. The probability $\pi_j$ case $j$ arising is given in the previous table. Note $c$ is the tail parameter of the underlying degree distribution. \label{tab:table2}}
\begin{ruledtabular}
\begin{tabular}{llll}
    $j$ & $\langle C_j \rangle$ & $ \langle C_j^2 \rangle - \langle C_j \rangle^2$ & $c_j$ \\
    \hline \hline
    1 & $\pi_1 (s\langle k \rangle + 1)$  & $\pi_1 (s^2 (\sigma^2 - \langle k \rangle + \langle k \rangle^2) + 2\langle k \rangle s) + \langle C_1 \rangle - \langle C_1 \rangle^2$ & $c \pi_1 s^{\gamma - 1}$       \\[1ex]
    2 & $\pi_2 s\langle k \rangle$        & $\pi_2 s^2 (\sigma^2 - \langle k \rangle + \langle k \rangle^2)         + \langle C_2 \rangle - \langle C_2 \rangle^2$ & $c \pi_2 s^{\gamma - 1}$       \\[1ex]
    3 & $\pi_3 (1 - s) \langle k \rangle$ & $\pi_3 (1 - s)^2 (\sigma^2 - \langle k \rangle + \langle k \rangle^2)   + \langle C_3 \rangle - \langle C_3 \rangle^2$ & $c \pi_3 (1 - s)^{\gamma - 1}$ \\[1ex]
    4 & $\pi_4 s\langle k \rangle$        & $\pi_4 s^2 (\sigma^2 - \langle k \rangle + \langle k \rangle^2)         + \langle C_4 \rangle - \langle C_4 \rangle^2$ & $c \pi_4 s^{\gamma - 1}$
\end{tabular}
\end{ruledtabular}
\end{table*}

To model the change in $s$, we denote by $C_j$ the random variable corresponding to change in $s$ at each update, described by $g_j(x)$, for each case $j=1, 2, 3$ and $4$ as above. After $n$ selections, a sequence $C_{j1}, C_{j2}, C_{j3}, \hdots, C_{jn}$ is produced, corresponding to the change in $s$ attributed to each case, with $C_{ji}$ being independent copies of $C_j$. In the following we shall use two forms of the central limit theorem~\cite{nolan2020univariate}. The first is the classic central limit theorem, which applies to sums of random variables whose variance is finite. The second is the generalised central limit theorem, which applies when the variance of the variables being summed is infinite. In both cases, the scaled and shifted sum of $n$ independent copies of $C_j$ converges in distribution to a random variable $Z$, such that 
\begin{equation}\label{eqn:clt}
  C_{j1} + C_{j2} + C_{j3} + \hdots + C_{jn} \xrightarrow[]{d} \dfrac{Z + b_{j, n}}{a_{j, n}}
\end{equation}
for constants $a_{j, n}$ and $b_{j, n}$, as $n \rightarrow \infty$. In the classic central limit theorem, $Z \sim \mathcal{N} (0, 1)$, the standard normal distribution, and in the generalised central limit theorem, $Z \sim \mathcal{S} (\alpha, \beta)$, with $\mathcal{S}$ the standardised stable distribution, and $\alpha$ and $\beta$ are shape parameters to be described in Sec.~\ref{sec:infinite}, as well as Appendix~\ref{app:stable}. As such, to approximate $\Delta s$, it remains only to relate the scale and shift constants, $a_{j, n}$ and $b_{j, n}$ for each case $j$ to the variable $s$, as well as the model parameters. The relevant results from the classic and generalised central limit theorems are given in Sec.~\ref{sec:finite} below.

Following the arguments above, the time evolution of $s$ can be found by repeatedly sampling random variables $Z$, and updating the constants $a_{j,n}$ and $b_{j,n}$ according to the central limit theorem at each step. Each step corresponds to $n$ updates, where the system time advances by $\Delta t$. This process can be expressed as a finite-difference regime, where
\begin{equation}\label{eqn:sdeclt}
    \Delta s = g(s) \Delta t + h(s) \Delta t^{1/2} Z, \,\text{ with } Z \sim \mathcal{N} (0, 1)
\end{equation}
and 
\begin{equation}\label{eqn:sdegclt}
    \Delta s = g(s) \Delta t + h(s) \Delta t^{1/\alpha} Z, \,\text{ with } Z  \sim \mathcal{S} (\alpha, \beta)
\end{equation}
when $p_k$ has finite- and infinite-variance, respectively. The drift and diffusion, $g(s)$ and $h(s)$, result from combining the terms $a_{j,n}$ and $b_{j,n}$, and are therefore functions of $s$, the innovation rate $\mu$, and the underlying network topology, as described below. We integrate Eqs.~(\ref{eqn:sdeclt}) and (\ref{eqn:sdegclt}) using discrete time steps (cf.~the Euler-Maruyama scheme~\cite{higham2001algorithmic}), producing trajectories of $s$ over time.

We briefly describe the initial condition on $s$ for Eqs.~(\ref{eqn:sdeclt}) and (\ref{eqn:sdegclt}). As described above, every meme enters the system through an innovation tweet, so it initially occupies the screen of the innovating node, as well as its $k$ followers. The initial value of $s$ is therefore $(1 + k) / N$, where $k$ is a random integer with distribution $p_k$. However, we find that setting $s(0) = (1 + \langle k \rangle) / N$ removes the dependence on $k$ and yields indistinguishable results, so we choose the latter in this work. Finally, we note that for finite $\Delta t$, fluctuations can cause the value of the density $s$ to decrease below zero, which is unphysical. As such, we treat $s = 0$ as an absorbing boundary condition, corresponding to the red meme going extinct.

\subsection{Finite-variance degree distribution}\label{sec:finite}

\begin{figure*}
    \centering
    \includegraphics[scale=1.0]{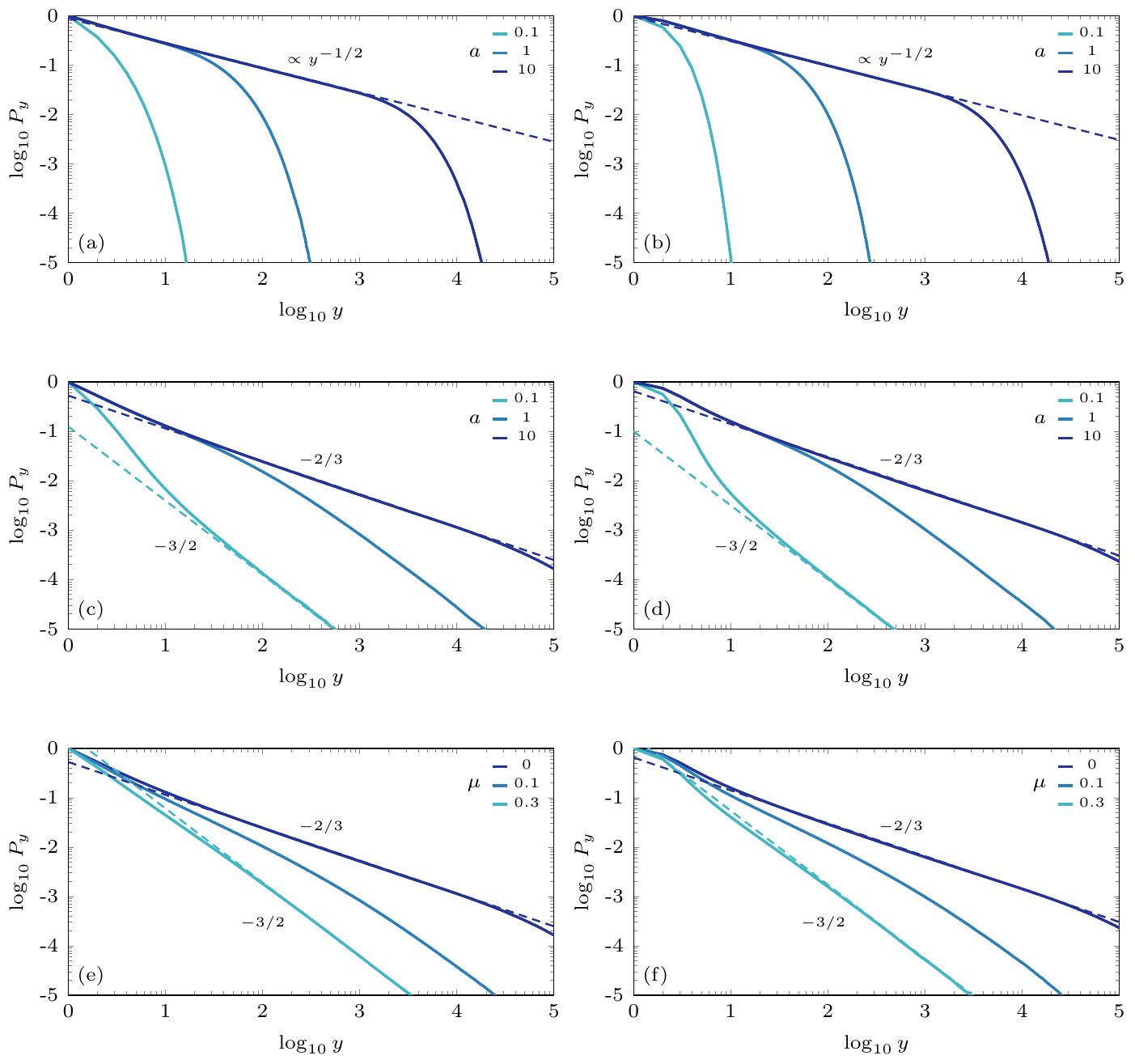}
    \caption{Comparison of the complementary cumulative distribution function of meme popularity $y$, $P_y$, for CIC simulations on networks, first column, and corresponding diffusion approximations, second column. Networks are of size $10^5$, and have 10-regular out-degree distribution in (a), and power-law out-degree distribution with exponent $\gamma = 2.5$ with lower bound $k = 4$ and resulting tail weight $c = 6.629$ in (c) and (e). Popularity distributions are measured for memes of age $a = 0.1, 1$ and $10$ with innovation probability $\mu = 0$ in (a) and (c), and $a = 10$ in (e), where $\mu$ is shown. Slopes on log-log scale are indicated, giving the exponent of the power-law fit. Panel (e) introduces nonzero innovation probability $\mu$, which is set to zero in (a) and (c). All diffusion approximation results set $n=10^3$. \label{fig:fig01}}
\end{figure*}

When the out-degree distribution $p_k$ of the underlying network has finite variance $\sigma^2$, fluctuations in the density $s$ of a given meme have finite variance. This is true on the whole, as well as for fluctuations of each type $j$. In each case, the sum of these fluctuations is approximated by $(Z_j + b_{j,n}) / a_{j,n}$, as per the classic central limit theorem, with the noise term $Z_j$ drawn from the standard normal distribution, $\mathcal{N}(0, 1)$. In order to compute $a_{j,n}$ and $b_{j,n}$, we require the mean and variance of the fluctuations of each type $C_j$. Both are straightforward to compute, as detailed in Table~\ref{tab:table2}. Then, $a_{j, n}$ is given by
\begin{equation}\label{eqn:ajnclt}
  a_{j, n} = \dfrac{1}{\sigma_j} n^{-\tfrac{1}{2}}
\end{equation}
and
\begin{equation}\label{eqn:bjnclt}
  b_{j,n} = n \langle C_{j} \rangle a_{j, n},
\end{equation}
where $\sigma_j^2 = \langle C_j^2 \rangle - \langle C_j \rangle^2$ is the variance of $C_j$, and $\langle C_j \rangle$ is its mean. We can then simplify the signed sum of the four $(Z_j + b_{j,n}) / a_{j,n}$ terms by grouping their deterministic and stochastic components, $b_{j, n} / a_{j, n}$ and $Z_j / a_{j, n}$ respectively. The former is simply $n\langle C_j \rangle$, and the latter can be simplified using rules for scaling and summing normal random variables. In this way, we produce the drift and diffusion terms in Eq.~(\ref{eqn:sdeclt}), respectively
\begin{equation}\label{eqn:gsclt}
    g(s) = -\langle C_1 \rangle - \langle C_2 \rangle + \langle C_3 \rangle - \langle C_4 \rangle,
\end{equation}
and
\begin{equation}\label{eqn:hsclt}
    h(s) = \left(\sigma_1^2 + \sigma_2^2 + \sigma_3^2 + \sigma_4^2 \right)^{\tfrac{1}{2}} N^{-\tfrac{1}{2}}.
\end{equation}
Recall that $\Delta t = n\, dt = n / N$, where $n$ is the number of update steps aggregated, and $N$ is system size. The $s$ dependency of $h(s)$ and $g(s)$ arises through $\langle C_j \rangle$ and $\sigma_j^2$, respectively, as per Table~\ref{tab:table2}. The noise term $Z$ associated with $h(s)$ in Eq.~(\ref{eqn:sdeclt}) is drawn from $\mathcal{N}(0, 1)$. The accuracy of this approximation can be seen in Fig.~\ref{fig:fig01}(a) and (b), and will be discussed below.

\subsection{Infinite-variance degree distribution}\label{sec:infinite}

In the case where the out-degree distribution $p_k$ has infinite variance, fluctuations in the density $s$ of a given meme have infinite variance. This is true on the whole, as well as for fluctuations $C_j$ of each type $j$. As such, the classic central limit theorem no longer applies. However, for some choices of $p_k$, namely, those whose tail probabilities are asymptotically  power-laws, with exponent $\gamma$ in the range $2 < \gamma < 3$, fluctuations of each type sum to $(Z_j + b_{j,n}) / a_{j,n}$, where $Z_j$ is an $\alpha$-stable random variable. This is described by the generalised central limit theorem \cite{nolan2020univariate}, with the variable $Z_j$ drawn from the standardised stable distribution $\mathcal{S}(\alpha, \beta)$, and $a_{j,n}$ and $b_{j,n}$ are norming constants (see Appendix~\ref{app:stable} for information regarding the parameterisation and sampling of stable distributions). 

The first shape parameter $\alpha$ is the stability parameter, and is related to the power-law decay of the variables being summed. To determine the value of $\alpha$ for fluctuations $C_j$ of each of the four types, we consider first the underlying degree distribution. In this work, we use power-law degree distributions $p_k \sim k^{-\gamma}$, where $2 < \gamma < 3$. As such, the sum of degrees themselves are asymptotically $\alpha$-stable, with $\alpha = \gamma - 1$. Since the change in the density of a given meme is closely related to the degree $k$ of the tweeting node, fluctuations also have tail probabilities decaying with power-law exponent $\gamma$. As a consequence, fluctuations $C_j$ of each type sum to $(Z_j + b_{j,n}) / a_{j,n}$ asymptotically, with $Z_j$ being $\alpha$-stable. This can be seen~\cite{faqeeh2019emergence} by examining the generating functions of these distributions, $g_j(x)$, in Table~\ref{tab:table1}.

The second shape parameter $\beta$ is related to skewness, and is determined by the relative weights of the upper and lower tail probabilities of the variable being summed \cite{nolan2020univariate}. The tail weights of fluctuations $C_j$ are related to the tail of the underlying degree distribution $p_k$. In particular, degrees have upper tail weight $c$ given by the relation $k^\alpha (1 - P_k) \rightarrow c$ as $k \rightarrow \infty$. Here $P_k$ is the cumulative distribution function of the degree distribution. In general for stable distributions, the lower tail weight is defined similarly, but here is zero since degrees are non-negative. In other words, $p_k$ is totally skewed to the right. The tail weights $c_j$ of the fluctuations $C_j$ are related to $c$, but moderated by the density $s$, as noted in the final column of Table~\ref{tab:table2}. Along with the stability parameter $\alpha$, the tail weights $c_j$ determine the scale term $a_{j,n}$ in the generalised central limit theorem, describing the sum of fluctuations of type $j$. For distributions of this type~\cite{nolan2020univariate}, the scaling constant $a_{j,n}$ in the generalised central limit theorem is given by
\begin{equation}
  a_{j, n} = \left( \dfrac{2\Gamma(\alpha)\sin (\tfrac{\pi \alpha}{2})}{\pi  c_j}\right)^{\tfrac{1}{\alpha}} n^{-\tfrac{1}{\alpha}},
\end{equation}
where $\Gamma$ is the gamma function, and $n$ as before is the number of terms in the sum, as per Eq.~(\ref{eqn:clt}), and the generalised central limit theorem. In the interval  $1 < \alpha < 2$ that is of interest, the value of $b_{j, n}$ is the same as in the finite-variance case, given by Eq.~(\ref{eqn:bjnclt}). In principle, computing $a_{j,n}$ and $b_{j,n}$, and sampling $\alpha$-stable variables (see Appendix~\ref{app:stable}) allows us to approximate the change in $s$ after $n$ node updates using Eq.~(\ref{eqn:sdegclt}).

As in the finite-variance case, we can simplify the signed sum of the four $(Z_j + b_{j,n}) / a_{j,n}$ terms by grouping together their deterministic and stochastic parts. First, the terms $b_{j,n} / a_{j,n}$ simplify as before to produce the drift $g(s)$, which is the same here as in Eq.~(\ref{eqn:gsclt}). Second, the signed sum over $Z_j / a_{j, n}$ can be computed using a basic property of stable laws, namely, that the sum of $\alpha$-stable random variables is also $\alpha$-stable. In this way, the diffusion term can be shown to be
\begin{equation}
    h(s) = \left(\dfrac{\pi(c_1 + c_2 + c_3 + c_4)}{2\Gamma (\alpha) \sin (\tfrac{\pi \alpha}{2})} \right)^{\tfrac{1}{\alpha}} N^{-\tfrac{\alpha - 1}{\alpha}},
\end{equation}
with the dependence of $h(s)$ on $s$ arising through the tail weights $c_j$, as per Table~\ref{tab:table2}. The random term $Z$ associated with $h(s)$ in Eq.~(\ref{eqn:sdegclt}) is drawn from the standardised stable distribution with shape parameters $\alpha$ and $\beta = (-c_1 - c_2 + c_3 - c_4) / (c_1 + c_2 + c_3 + c_4)$. Here we note that while the distributions of $C_j$ are totally skewed to the right, their signed sum is not. That is, since $j = 1, 2$ and $4$ contribute negatively, meaning the lower tail is no longer zero. The $j = 3$ case contributes positively, giving the upper tail. As such, the value of $\beta$ for the signed sum of the terms $Z_j / a_{j,n}$ is given by the relative weights of the upper and lower tails.

Finally, it is useful to point out some differences in Eqs.~(\ref{eqn:sdeclt}) and (\ref{eqn:sdegclt}), particularly in regards to the factor $\Delta t = n\, dt = n / N$ in the diffusion term. In the finite-variance case noise $Z$ scales as $\Delta t^{1/2}$, and in the infinite-variance case scales as $\Delta t^{1 / \alpha}$. When $\alpha = 2$, stable theory recovers the normal case. Further, it may be noteworthy that $Z$ is independent of $s$ in Eq.~(\ref{eqn:sdeclt}), but dependent on $s$ in Eq.~(\ref{eqn:sdegclt}), via the skewness parameter $\beta$.

\subsection{Popularity}

In the dynamics of the model introduced in Ref.~\cite{gleeson2014competition}, we associate with a meme not only a time-dependent screen density $s$, but also a popularity, $y$, a non-negative integer indicating the number of times a given meme has been retweeted since it first appeared. Because popularity is observable in online social networks like Twitter, where it corresponds to retweets, it is easier to compare with data than screen abundance. As such, we model the evolution of $y$ along with $s$. The is straightforward to do, since $y$ is simply incremented by one when a selected node contains the red meme and retweets rather than innovates. This describes case $j = 3$, which occurs with probability $\pi_3$, as per Table~\ref{tab:table1}. We model the fluctuations in $y$ over $n$ node selections as $n$ samples from the Bernoulli distribution with probability $\pi_3$, resulting in the equation
\begin{equation}\label{eqn:popularity}
    \Delta y = g(s)\Delta t + h(s) \Delta t Z, \,\text{ with }Z\sim\mathcal{N}(0, 1),
\end{equation}
where $g(s) = \pi_3 N$, the expected value of the Bernoulli distribution, and $h^2(s) = \pi_3 (1 - \pi_3) N$, the standard deviation of the same distribution. Finally, note that in simulation there is a boundary condition, where $y$ must remain constant if $s$ goes to zero. Note also that $y$ in Eq.~(\ref{eqn:popularity}) is continuous due to the noise term, so we round $y$ to the nearest integer when reporting it below.

\section{Results}\label{sec:results}

Here we discuss a series of computational experiments in which we validate the diffusion approximation through comparison to network-based simulation of the CIC model. We begin by surveying a range of model parameters in Section~\ref{sec:comparison} to verify that the diffusion approximation is effective in a variety of settings. Then in Section~\ref{sec:fp} we use the Fokker-Planck equation to derive an expression of the time evolution of the moments of the density $s$ of a given meme. Finally in Section~\ref{sec:accuracy} we examine the effect of the time step $\Delta t = n\, dt$ on the accuracy of the diffusion approximation.

\subsection{Accuracy of diffusion approximation}\label{sec:comparison}

In Fig.~\ref{fig:fig01} we use the diffusion approximation to recover a number of experimental results observed in network-based simulation. In Fig.~\ref{fig:fig01}(a), we measure the distribution of popularities $y$, as a function of meme age $a$, which is the time elapsed between when a meme was introduced to the system through innovation, and observation time $t$. The follower distribution $p_k$ is 10-regular, and the innovation rate $\mu$ is zero, except initially when blank screens are selected. We observe that at all the  ages shown, $a = 0.1, 1$ and $10$, the tail of complementary cumulative distribution of popularities $P_y$ decays exponentially. For larger ages, popularities follow a power law with exponent $-1/2$. This is typical of avalanche size distributions at a dynamical critical point, or of outbreak sizes in a neutral model \cite{pinto2011quasi}, and recovers the results of Ref.~\cite{gleeson2014competition}. These observations are reproduced almost exactly by the diffusion approximation, see Fig.~\ref{fig:fig01}(b), particularly when $a$ is large. When $a = 0.1$, there appears to be a very slight difference in the rate of the exponential cutoff. The diffusion approximation uses $n = 10^3$.

In Fig.~\ref{fig:fig01}(c), we repeat this experiment almost identically, the only difference being that the follower distribution $p_k$ is power-law, rather than regular. The power-law exponent is $\gamma = 2.5$, and we set a minimum degree cutoff of $k = 4$. Here we recover a well-known result, explained in Ref.~\cite{gleeson2014competition}, that popularity $P_y$ decays as a power-law with exponent $-2/3$, for large ages $a$. For early ages, a power law decay is also apparent, but this appears to be a reflection of the out-degree distribution. That is, when a meme first enters the system, it appears on the screen of the innovating node, as well as its $k$ followers. Since the follower distribution has exponent $-5/2$, this appears as $-3/2$ in the ccdf picture. Once again, these observations are recovered with high accuracy by the diffusion approximation in Fig.~\ref{fig:fig01}(d), especially for larger $a$.

Finally, in Fig.~\ref{fig:fig01}(e) we examine the effect of allowing innovation with probability $\mu = 0.1$ and $0.3$. We restrict ourselves to the popularity distribution for memes of age $a = 10$. Otherwise, experiments are identical to Fig.~\ref{fig:fig01}(c), which is reproduced here as the $\mu = 0$ case for comparison. When $\mu = 0.1$, the distribution of popularities $y$ is made more narrow. For larger values, $\mu = 0.3$, the distribution once again tends to a power law with exponent $-3/2$. As before, this appears to be a sign that there are relatively few memes to survive to this age, due to the increased competition for screen space. As such, most memes die out soon after being born, meaning their popularity was proportional to the number of followers $k$ that they were initially broadcast to. This is supported by the fact that this appears to be a limiting distribution for $\mu > 0.3$, not shown. These results are supported by the diffusion approximation, Fig.~\ref{fig:fig01}(f).

\subsection{Fokker-Planck equation and its moments}\label{sec:fp}

\begin{figure}
    \hspace*{-4mm}
    \includegraphics[scale=1.0]{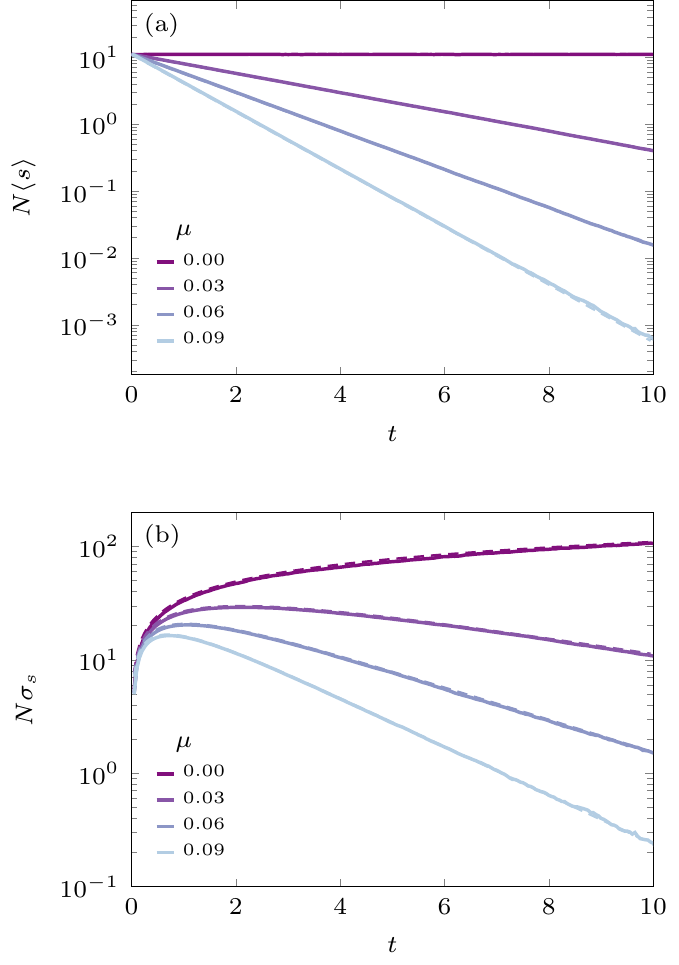}
    \caption{Mean $\langle s \rangle$ and standard deviation $\sigma_s$, (a) and (b) respectively, of the screen density $s$ of a given meme, as a function of the time $t$ since the meme first appears in the system. Solid curves result from simulation of networks of size $N = 10^6$, for a duration of $100$ macroscopic time steps. Dashed curves are the results from the Fokker-Planck equation with linearised drift and diffusion terms, given by Eqs.~(\ref{eqn:fpmean}) and (\ref{eqn:fpvar}). The follower distribution $p_k$ is $10$-random regular. \label{fig:fig03}}
\end{figure}

In this section we use the Fokker-Planck equation to derive the mean and variance of meme density $s$, the fraction of screens occupied by a given meme, as a function of the time since it was introduced. It is this that we refer to by time $t$ from now on, rather than total system time. We restrict ourselves to the case where the follower distribution $p_k$ has finite-variance. The Fokker-Plank equation~\cite{sornette2006critical} describes the time evolution of $p(s, t)$, the probability density function of screen fraction $s$ as a function of time $t$. Beginning with a Langevin equation describing a single trajectory, such as Eq.~(\ref{eqn:sdeclt}), the Fokker-Planck equation can be written 
\begin{equation}\label{eqn:fp}
  \dfrac{\partial}{\partial t}p(s,t) = -\dfrac{\partial}{\partial s}[g(s) p(s,t)] + \dfrac{\partial^2}{\partial s^2}[ D(s) p(s,t)],
\end{equation}
where $g(s)$ and $D(s) = h^2(s) / 2$ are the drift and diffusion terms of the single particle equation, defined in Eqs.~(\ref{eqn:gsclt}) and (\ref{eqn:hsclt}), respectively. We assume that initially, $p(s,0) = \delta((1 + \langle k \rangle) / N)$, where $\delta$ is the Dirac delta, since all individual trajectories $s$ start at this value. As it stands, the Fokker-Planck equation cannot be solved explicitly for our system, but can of course be integrated numerically using standard techniques.

It is possible to obtain an analytic estimate for $\langle s \rangle$ and $\sigma_s$, without solving the Fokker-Planck equation for $p(s,t)$ directly. To this end, observe that at time $t$, the $m$-th moment of $s$ is given by
\begin{equation}
    \langle s^m \rangle = \int_0^1 s^m p(s,t) ds.
\end{equation}
As such, by multiplying both sides of Eq.~(\ref{eqn:fp}) by $s^m$, and integrating over all $s$, we obtain an ordinary differential equation describing the time evolution of the $m$th moment $\langle s^m \rangle$. In general, this too cannot be solved in closed form. However, by including only the first-order terms in $s$ in the drift and diffusion expressions $g(s)$ and $h(s)$, respectively, the equation becomes tractable. In particular, when $m = 1$ and $2$, the resulting equation can be solved for $\langle s \rangle$ and $\langle s^2 \rangle$ using standard techniques, which in turn yield an expression for $\sigma_s$. Concretely, this procedure yields 
\begin{equation}\label{eqn:fpmean}
    \langle s \rangle = \kappa_1 e^{-\mu(\langle k \rangle + 1)t}
\end{equation}
and
\begin{equation}\label{eqn:fpvar}
    \sigma_s^2 = \kappa_2 \left( e^{-\mu(\langle k \rangle + 1)t} - e^{-2\mu(\langle k \rangle + 1)t} \right),
\end{equation}
where $\kappa_1$ and $\kappa_2$ are constant functions of network topology, through $\langle k \rangle$ and $\sigma$, and the innovation probability $\mu$. The first constant, $\kappa_1 = (\langle k \rangle + 1) / N$, is simply the initial value of $s$, while the second, $\kappa_2 = (\mu + \langle k \rangle + (\langle k \rangle^2 + \sigma^2)(1 - \mu)) / \mu N^2$, indicates that $\sigma_s$ is an increasing function of $\langle k \rangle$ and $\sigma$. The effectiveness of this approach is evident in Fig.~\ref{fig:fig03}, where it is clear that the first-order approximation of $s$ is appropriate. In this figure, solid curves result from network-based simulation of CIC. For each value of $\mu$, we simulate the network dynamics for $100N$ update steps, and track the screen count $s$ for each meme, as a function of the time since it was introduced to the system. There appears to be only a very small systematic error in $\sigma_S$, with the theoretical moments overestimating the value observed in network simulation. It is likely that the linearisation of the diffusion term causes this. No such error is apparent in Fig.~\ref{fig:fig03}(a).

\subsection{Dependence on $n$, or time step $\Delta t$}\label{sec:accuracy}

In this section we examine how the parameter $n$ affects the quality of the diffusion approximation. This parameter is defined in Eq.~(\ref{eqn:clt}), and is the number of update events that are aggregated to produce a single time step of the diffusion approximation. It appears in the finite-difference implementation, Eqs.~(\ref{eqn:sdeclt}) and (\ref{eqn:sdegclt}), as $\Delta t = n\, dt = n / N$. Our choice of $n$ in practice is determined by two competing effects. On one hand, the larger the value of $n$, the better the approximation of the sum of the fluctuations $C_j$ as a normal distribution, as per the central limit theorem. This is especially true when $s$ is small, such as at early times. On the other hand, the smaller the value of $n$, the better the finite-difference solution. It is generically true that finite-difference methods are more accurate the smaller the value of $\Delta t = n\, dt$, as we assume that the variable in question, here $s$, is constant over the integration time step. As such, $n$ must be large enough for the central limit theorem to apply, but small enough that $s$ does not vary appreciably over that interval.

\begin{figure}
    \hspace*{-4mm}
    \includegraphics[scale=1.0]{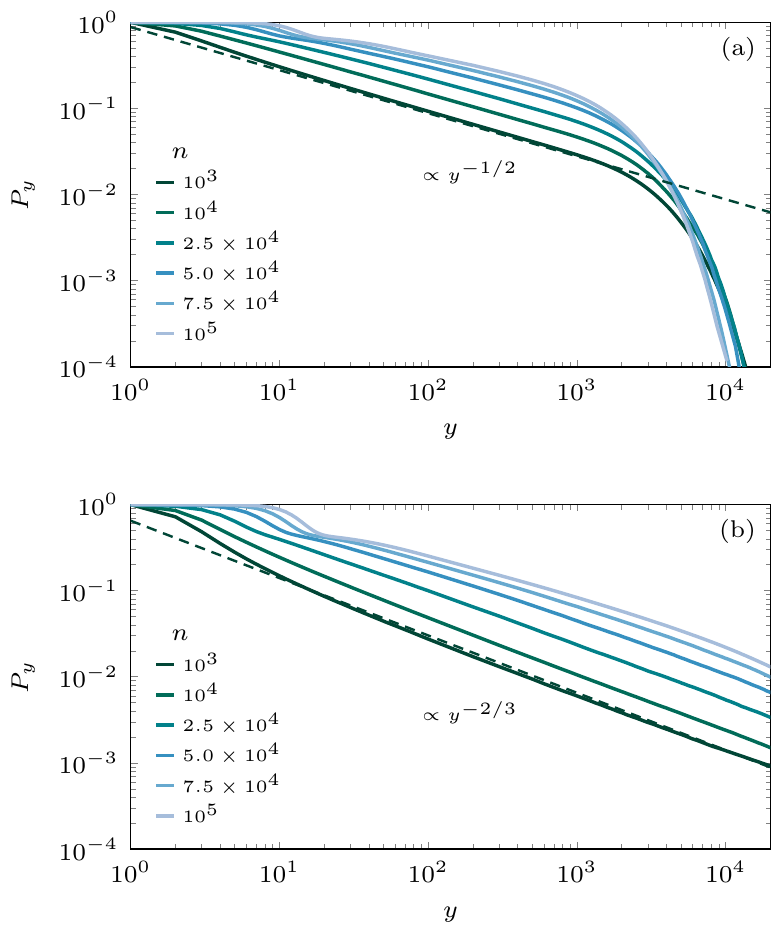}
    \caption{Complementary cumulative distribution of meme popularity $y$, using the diffusion approximation with various time step sizes $\Delta t = n\, dt = n / N$. The follower distribution $p_k$ is $10$-random regular, (a), and power-law with lower cutoff at $k = 4$, exponent $2.5$, and resulting tail weight $c = 6.629$, (b). For each value of $n$, $10^7$ trajectories were computed, either until memes reached an age of $a = 10$, or until they died out. Curves at $n = 10^3$ in (a) and (b) correspond to the $a = 10$ curves in Fig.~\ref{fig:fig01}(b) and (d), respectively. \label{fig:fig02}}
\end{figure}

We verify the effect of increasing $n$ in Fig.~\ref{fig:fig02}(a) and (b), where solid curves show the result of the diffusion approximation. Experimentally, we use the same settings as in Fig.~\ref{fig:fig01}(b)
for memes at ages $a = 10$. We set $n = 10^3$ in that figure, which is reproduced as a baseline result here because it matches well to the numerical results of Fig.~\ref{fig:fig01}(a). The popularity curves are visibly distorted as we increase $n$ from $10^3$ up to $10^5$, both when the follower distribution $p_k$ is regular and power-law. It is difficult to know \emph{a priori} which value of $n$ to choose, in this work we have used a value of $n = 10^3$, unless stated otherwise.

\section{Discussion}

In this work we have approximated a model of meme popularity as a one-dimensional stochastic process, and shown it to be effective in recovering a range of critical spreading phenomena on networks. Depending on whether the degree distribution has finite or infinite variance, dynamic fluctuations can be approximated by Gaussian and $\alpha$-stable noise, respectively. The effectiveness of this approximation is revealing, shedding light on the mechanisms that ultimately drive dynamical outcomes. In particular, our results show that in the finite-variance case, only the first two moments $\langle k \rangle$ and $\sigma$ of the follower distribution $p_k$ play a role. In the infinite variance case, only the mean $\langle k \rangle$, the power-law exponent $\alpha$ and its tail weight $c$ are important. As such, any approximation using the full out-degree distribution $p_k$ is wasteful by comparison.
In addition, the diffusion approximation shows that memes can be modelled independently of one another. That is, each realisation of the Langevin equations, modelling the spread of a single meme, is independent of all other realisations. A clear advantage of this is that trajectories can be computed in parallel, something that is impossible in network-based simulations of the dynamics. This gain in efficiency is especially noteworthy in the limit of large networks, where memory is obviously a constraint when simulating networks themselves, in contrast to the diffusion approximation. In a similar vein, the diffusion approximation allows us to sample as many, or as few trajectories as we like. While a large number of memes may emerge in a large network, their distribution may be revealed by a much smaller sample of memes using the diffusion approximation. Small networks may present the opposite problem, in which case the diffusion approximation allows us to extrapolate by sampling additional trajectories.

Another advantage of the diffusion approximation is that it naturally gives rise to a suite of analytical tools. In particular, diffusion processes have a corresponding Fokker-Planck equation, which we have exploited to derive expressions for the temporal evolution of the moments of meme abundance as an ensemble. We used this to calculate the mean and variance of meme abundance as a function of time and demonstrated that a small-$s$ linearisation gives accurate closed-form solutions. This new perspective paves the way for future work, given the ease with which slight modifications can be made to allow the study of different meme spreading dynamics, such as when memes attract user attention  heterogeneously (i.e., some are more likely to spread than others), sometimes referred to as fitness models~\cite{oliveira2019effects}.

One limitation of our work is that we have assumed that the in-degree, meaning the number of individuals that a give node follows, is constant across the network. Allowing this distribution to be arbitrary leads to small, but noticeable disagreements in our results, such as the accuracy of the Fokker-Planck approximation of the moments of $s$. Extending our analytical results to account for this is a potential direction for future work.

\section*{Acknowledgements}
This work was partly supported by Science Foundation Ireland under grant numbers 16/IA/4470 (all authors), 12/RC/2289 P2 (J.G.) and 16/RC/3918 (J.G.).
\\
\leqnomode
\begin{flalign*}
    & \quad  \text{Corresponding author:} \textit{ Kleber.Oliveira@ul.ie} \label{corresp} \tag*{$^\dagger$} &
\end{flalign*}
\reqnomode
\appendix

\section{Sampling stable variables}\label{app:stable}

In this appendix we describe the numerical procedure for simulating $\alpha$-stable variables as is necessary for the diffusion term in Eq.~(\ref{eqn:sdegclt}). Following the formulation of the Chamber-Mallows-Stuck algorithm from Ref.~\cite{nolan2020univariate}, we let $\Theta$ and $W$ be independent random variables with $\Theta$ uniformly distributed on $(-\tfrac{\pi}{2}, \tfrac{\pi}{2})$ and $W$ exponentially distributed with mean $1$. We focus on the interval $1 < \alpha < 2$. For the skewness parameter $\beta$ in Eq.~(\ref{eqn:sdegclt}), we have $-1 \leq \beta \leq 1$, determined by the tail parameters of the variables being summed, as described in the main text. We define $\theta = \tan^{-1} (\beta \tan (\pi \alpha / 2)) / \alpha$. Then, the random variable $Z$ given by
\begin{equation}
    Z \sim \dfrac{\sin \alpha (\theta + \Theta)}{(\cos \alpha \theta \cos \Theta)^{1/\alpha}} \left(\dfrac{\cos ((\alpha - 1)\Theta)}{W} \right)^{\tfrac{1-\alpha}{\alpha}}
\end{equation}
follows a $\mathcal{S}(\alpha, \beta)$ distribution~\cite{nolan2020univariate}. In this paper we have used $\alpha = 1.5$ throughout.  For clarity, recall that $\alpha$ here is related to the exponent of the power-law degree distribution $\gamma$ by $\alpha = \gamma - 1$, and $\beta$ is determined by the relative weights of the tail probabilities $c_j$, specifically $\beta = (-c_1 - c_2 + c_3 - c_4) / (c_1 + c_2 + c_3 + c_4)$, which in turn are defined in Table~\ref{tab:table2}. In the same vein, normal variables can be obtained starting with uniform random variables using the Box-Muller transform. Note that two parameterisations are outlined in Ref.~\cite{nolan2020univariate}, and flagged notationally by 0 or 1. Throughout this work we adhere to the latter, which is written $\mathcal{S}(\alpha, \beta; 1)$ in Ref.~\cite{nolan2020univariate}.



\bibliography{apssamp}

\end{document}